# CURRICULUM BASED MULTI-TASK LEARNING FOR PARKINSON'S DISEASE DETECTION

*Nikhil J. Dhinagar\*[1], Conor Owens-Walton[1], Emily Laltoo[1], Christina P. Boyle[1],
Yao-Liang Chen[2], Philip Cook[3], Corey McMillan[3], Chih-Chien Tsai[4],
J-J Wang[5], Yih-Ru Wu[6], Ysbrand van der Werf[7], Paul M. Thompson[1]*

[1]Imaging Genetics Center, Mark & Mary Stevens Institute for Neuroimaging & Informatics,
Keck School of Medicine, University of Southern California, Los Angeles, CA, USA
[2]Department of Diagnostic Radiology, Chang Gung Memorial Hospital, Keelung, Taiwan
[3]Department of Neurology, University of Pennsylvania Perelman School of Medicine,
Philadelphia, PA, USA
[4]Healthy Aging Research Center, Chang Gung University, Taoyuan, Taiwan
[5]Department of Medical Imaging and Radiological Sciences, Chang Gung University, Taoyuan, Taiwan
[6]Department of Neurology, Chang Gung Memorial Hospital, Linkou, Taoyuan, Taiwan
[7]Amsterdam UMC, Vrije Universiteit Amsterdam, Department of Anatomy & Neurosciences, Amsterdam Neuroscience,
Amsterdam, The Netherlands.

**ABSTRACT**

There is great interest in developing radiological classifiers for diagnosis, staging, and predictive modeling in progressive diseases such as Parkinson's disease (PD), a neurodegenerative disease that is difficult to detect in its early stages. Here we leverage severity-based meta-data on the stages of disease to define a curriculum for training a deep convolutional neural network (CNN). Typically, deep learning networks are trained by randomly selecting samples in each mini-batch. By contrast, curriculum learning is a training strategy that aims to boost classifier performance by starting with examples that are easier to classify. Here we define a curriculum to progressively increase the difficulty of the training data corresponding to the Hoehn and Yahr (H&Y) staging system for PD (total N=1,012; 653 PD patients, 359 controls; age range: 20.0-84.9 years). Even with our multi-task setting using pre-trained CNNs and transfer learning, PD classification based on T1-weighted (T1-w) MRI was challenging (ROC AUC: 0.59-0.65), but curriculum training boosted performance (by 3.9%) compared to our baseline model. Future work with multimodal imaging may further boost performance.

## 1. INTRODUCTION

According to the World Health Organization (WHO), PD affects around 8.5 million people worldwide, and is one the most common neurodegenerative diseases globally [1]. PD usually affects older individuals and involves motor symptoms including tremor, rigidity, and difficulty walking often accompanied by depression, sleep problems, and cognitive decline [2]. Though PD is diagnosed clinically, there is interest in whether deep learning applied to radiologic images, such as standard T1-w brain MRIs, could be useful for PD detection, staging, and subtyping, and to identify characteristic brain biomarkers of PD.

Large-scale multicenter analyses of PD have found subtle but robust differences on brain MRI [3] and in other imaging modalities such as diffusion tensor imaging (DTI) [4]. Pooling brain MRI and clinical data from 2,357 PD patients and 1,182 healthy controls, from 19 sources worldwide, Laansma et al. [3] found subtle brain differences in stage 1 of PD, with volumetric deficits initially in temporal, parietal, and occipital cortices, spreading to the putamen, amygdala, and rostrally located cortical regions with increasing disease severity. From PD stage 2 onward, the bilateral putamen and amygdala were consistently smaller with greater deficits at later stages, but differences were subtle at stage 1 of PD. The subtle differences in early PD have made disease classification much more challenging than in Alzheimer's disease (AD), for example, with typical ROC-AUCs of 0.8 or greater for AD and 0.6-0.7 for PD, depending on the stage of PD and the method used [5].

Soltaninejad et al. used regional brain morphometry Logistic Regression (LR), Random Forest (RF) and Support Vector Machine (SVM), with machine learning classifiers for PD classification [6]. Tested on the PPMI (Parkinson's Progressive Markers Initiative) dataset only, they achieved an ROC-AUC of 0.5-0.75 for various models. RF models performed the best, but they did not test the model across multiple datasets, which can lower performance due to domain shift (differences in scanners and populations studied). In a recent dissertation, E. Yagis used 2D and 3D CNNs for PD classification in the PPMI dataset, attaining 61-67% accuracy depending on the architecture [7]. In a recent innovation, Zhao et al. [8] fitted an ordinal regression model to surface-based 3D meshes extracted from brain MRI, where local metrics from the basal ganglia and other subcortical surfaces were used to stage the disease. Using diffusion MRI, Zhao et al. [9] obtained promising PD classification performance with a CNN trained and tested on data from 305 PD patients and 227 healthy controls, suggesting the value of imaging data beyond standard T1-w MRI. Still, interest remains in understanding how well T1-w MRI can classify PD, as it is more routinely collected than diffusion MRI.

In this work, we used the H&Y stages as meta-data for T1-w MRI images via curriculum learning. Bengio et al. presented one of the earlier works on the benefits of curriculum learning on the training process [10]. Recent works have shown the promise of curriculum learning for different applications [11][12][13]. Wei et al. showed a performance improvement with a curriculum strategy for pathology image classification using annotator agreement as a proxy for example difficulty [14].

To this end, we propose a curriculum-based multi-task framework to classify PD using only structural T1-w MRI images, that leverages the H&Y staging system in the training phase. Our main contributions are:
1. We used clinically relevant information, i.e., H&Y staging, to define a curriculum for PD classification; and
2. We also used a multi-task approach to generate a larger set of re-usable features for detection of PD.

## 2. DATA
### 2.1 Neuroimaging Datasets
We analyzed 3D T1-w brain MRI and clinical data from three independent cohorts, summarized in **Table 1**. The cohorts were from: (1) Chang Gung University, Taiwan (467 scans); (2) the University of Pennsylvania (UPenn; 164 scans), and (3) the Parkinson's Progression Markers Initiative (PPMI; 381 scans), a multisite international study of PD. We included all T1-w scans from the cohorts that had a valid image file, diagnosis, and complete meta-data including age, sex, and H&Y staging information. **Fig. 1** visualizes staging information in proportion to the size of the three cohorts. We partitioned the Taiwan dataset in a ratio of 80:10:10 to create training, validation, and test sets. The UPenn and PPMI cohorts were used as out-of-distribution (OOD) zero-shot test datasets.

### 2.2 3D T1-weighted MRI Pre-processing
In line with similar studies [15], all T1-w brain MRI scans were pre-processed via standard steps for neuroimaging analyses including: nonparametric intensity normalization (N4 bias field correction) [16], 'skull-stripping', linear registration to a template with 9 degrees of freedom, and isotropic resampling of voxels to 2-mm resolution. The input spatial dimension of the MRIs was 91x109x91. All images were $z$-transformed (setting each image's mean and SD to a standard value) to stabilize model training.

**Table. 1.** Parkinson's Disease data analyzed in this paper.

| | Training (N = 378) | Validation (N = 42) | Test (N = 47) | OOD Test 1 (N = 164) | OOD Test 2 (N = 381) |
|---|---|---|---|---|---|
| Age Range (years) and average [SD] | 20.0-80.0 (60.69 [8.75]) | 39.0-80.0 (60.97 [9.18]) | 38.0-79.0 (60.06 [8.44]) | 50.0-86.0 (66.79 [7.72]) | 30.6 -84.9 (61.3 [9.9]) |
| Female | 194 (51%) | 20 (48%) | 22 (47%) | 63 (38%) | 133 (35%) |
| Parkinson's Disease Patients | 198 (52%) | 21 (50%) | 25 (53%) | 133 (81%) | 276 (72%) |
| Staging (stages 0 through 4 + controls) | s4-27; s3-43; s2- 67; s1-61; s0–0; CN -180 | s4-3; s3-5; s2-7; s1-7; s0–0; CN-20 | s4-3; s3 - 5; s2- 9; s1 - 8; s0 – 0; CN – 22 | s4-8; s3-72; s2-46; s1-7; s0–0; CN–31 | s4-0; s3-1; s2-162; s1-114; s0–0; CN – 104 |
| Sites: | Taiwan | Taiwan | Taiwan | UPenn | PPMI |

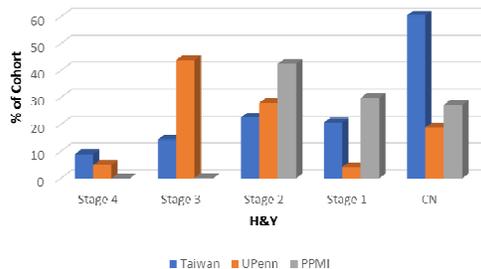

**Fig. 1.** Visualization of H & Y stages for Taiwan, UPenn, PPMI datasets. Stages are shown in the reverse of the usual order - to emphasize that we train the classifier on the most severely affected patients first (i.e., 'curriculum learning').

## 3. METHODS
### 3.1 Feature Extraction Backbone
We used a 3D CNN version of the DenseNet121 architecture [17]. This architecture has been effective in various neuroimaging applications [18][19]. We also used a scaled down version of the DenseNet121 called the Tiny-DenseNet [20] which was designed to reduce the number of parameters that need to be trained by a factor of 10. Transfer learning offers an effective way to re-use and adapt domain-specific features for new tasks [9]. We used pre-training as an initialization for our backbones. We pre-trained our backbone on 37,176 T1-w brain MRI scans from the UK Biobank with supervised learning, using sex-classification [21] as the pre-training task.

### 3.2 Curriculum Learning
The pre-trained backbone provides a model with relevant features learned *a priori* that can be further fine-tuned via curriculum learning for the PD classification task. We defined our *curriculum* in an episodic manner [14] by means of the disease severity staging information. In each progressive episode, we sample data from a lower H&Y stage to fine-tune the CNN, iteratively. For fine-tuning, episode 1 includes controls + stage4, episode 2 includes controls + stage4 + stage3, episode 3 includes controls + stage4 + stage3 + stage2, and episode 4 includes controls + stage4 + stage3 + stage2 + stage1. We overlap the stages in each episode to prevent 'catastrophic forgetting'. This framework is illustrated in **Fig. 2**. The intuition behind this design is that greater brain abnormalities are typically evident in more advanced stages of PD, making the relevant brain features easier to learn. We tested an *anti-curriculum* - the inverse of the *curriculum* defined earlier - i.e., training starts from stage 1 and works its way up to stage 4. We also tested the effect of balancing the number of controls with the number of PD cases in each episode, as the availability of patient data with higher stages of PD was limited.

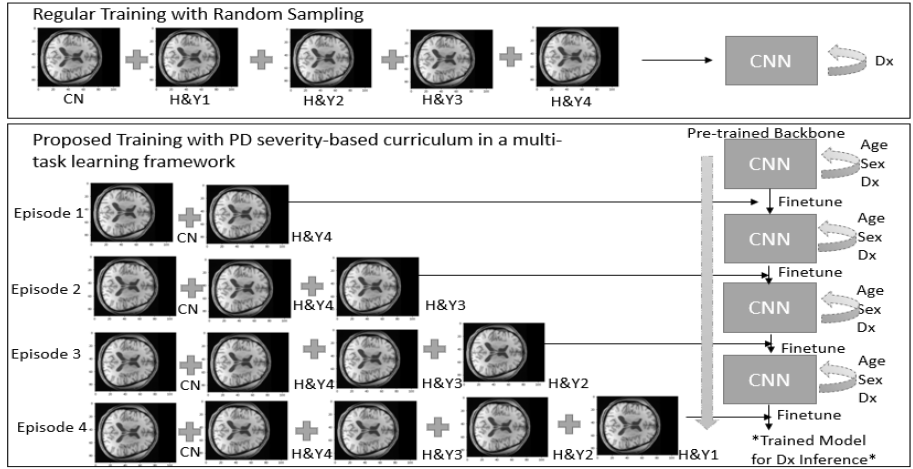

**Fig. 2. Proposed Training with a PD severity-based curriculum in a multi-task learning framework.** Regular training typically involves training the model on randomly ordered data. In this work, we propose the use of multi-task learning, implemented in the context of defining a curriculum - here based on H&Y severity scores.

### 3.3 Multi-Task Learning

Next, we implemented a *multi-task learning* paradigm to favor the learning of generally useful features relevant for the downstream task. Caruana [22] described *multi-task learning* as a way to "improve generalization by leveraging domain-specific information". In each training episode of our framework, the model was trained to predict the age, sex, and diagnosis of the sample in the mini-batch. We used our multi-task learning approach to include prediction of age and sex, since both variables are well-known confounding factors with the associated neuroimaging data. This was achieved by creating three separate task-specific heads that use features from a common feature backbone. The losses for each of the three tasks were summed to obtain the *total loss* per mini-batch:

$$L_T = L_{age} + L_{sex} + L_{dx} \quad (1)$$

Here, $L_{age}$ is the $L1$ loss = |real age – predicted age| (summed over the minibatch), and $L_{sex}$ and $L_{dx}$ are the binary cross-entropy loss. The L1 loss was used for the age regression task, to avoid a large discrepancy in the scales of the various losses. The classification tasks used a standard binary cross-entropy loss function.

### 3.4 Hyperparameter Optimization and Model Training

The Kaiming initialization scheme was used to initialize our CNNs [23]. We performed a random search to select key hyperparameter values, including the learning rate {2e-3 to 1e-5}, optimizer {sgd, ADAM, ADAM with weight decay}, and batch size {1, 4, 8, 16}. The model was trained for 30 epochs per episode within our proposed framework. We stopped training if there was no improvement in the validation loss for 15 to 20 epochs. We used a batch size of 16 except for the Tiny-DenseNet, where we used 8. The CNNs were trained with a learning rate of 0.0003 with the ADAM [24] optimizer except for using 0.0009 and 0.0001 with SGD (stochastic gradient descent) while training the Tiny-DenseNet and the DenseNet121 from scratch.

### 3.5 Model Testing

We evaluated the model with a set of performance measures including the receiver-operator characteristic curve-area under the curve (ROC-AUC), and the accuracy, precision based on a threshold optimized using Youden's Index [25]. We present in our results the average over subsequent runs of the model for each experiment and its standard deviation. We test our models on two out-of-distribution test sets without any additional fine-tuning.

### 3.5 Model Interpretation

Though several methods have been used in prior works to facilitate model interpretation, sometimes it is difficult to interpret these interpretability approaches. In this paper we conducted an intuitive occlusion sensitivity [26] analysis to visualize the localized regions focused on by the model. This involves systematically masking regions of the image with a black patch and aggregating the CNN predictions to create a heatmap. We empirically selected a patch size of 16 and stride of 4 for our experiments.

## 4. RESULTS

We were able to boost performance for PD classification by defining a *curriculum* based on the H&Y stages to fine-tune our pre-trained backbone CNN in a *multi-task learning* framework Our proposed approach (DenseNet121 with Curriculum + Multitask) achieved an average ROC-AUC of 0.653 (0.005) compared to our baseline (DenseNet121) of 0.604 (0.011) when our 3D CNN was trained from scratch. Our approach achieved an average test accuracy and precision of 0.674 (0.009) and 0.684 (0.011) compared to 0.653 (0.010), 0.628 (0.007) respectively for the baseline. We also experimented with using balanced subsets within each episode of the *curriculum* and an *anti-curriculum* to record the effect on the training. We present these results in **Table 2**. Our pre-training involves supervised learning on the UKB dataset with 37,176 MRI scans, referred to as UKB37K SL in the Table. We also show in **Table 2** the independent model evaluation on two out-of-distribution (OOD) datasets, i.e., UPenn and PPMI, without any additional fine-tuning.

**Table 2.** Parkinson's disease classification with benchmarks using curriculum learning and multi-task learning.

| Architecture | Pre trained Backbone | Fine-tuning Strategy | Taiwan Test ROC-AUC | UPenn Test zero-shot ROC-AUC | PPMI Test zero-shot ROC-AUC |
|---|---|---|---|---|---|
| DenseNet121 | UKB37K SL | Curriculum +Multitask | **0.653 (0.005)** | 0.541 (0.011) | 0.568 (0.004) |
| DenseNet121 | UKB37K SL | Curriculum (balanced) +Multitask | 0.646 (0.009) | 0.501 (0.011) | 0.567 (0.001) |
| DenseNet121 | UKB37K SL | Anti-Curriculum +Multitask | 0.626 (0.005) | 0.542 (0.006) | 0.528 (0.008 |
| DenseNet121 | - | - | 0.614 (0.013) | 0.582 (0.009) | 0.552 (0.001) |
| DenseNet121 | UKB37K SL | Anti-Curriculum (balanced) +Multitask | 0.604 (0.011) | 0.569 (0.007) | 0.525 (0.003) |
| TinyDenseNet | - | - | 0.590 (0.002) | 0.571 (0.007) | 0.554 (0.015) |

We illustrate the occlusion sensitivity analysis in **Fig. 3**. We create heat maps based on our approach for scans at different H&Y stages that are superimposed on the scans. The figure shows the map for the CNN trained with and without curriculum learning.

## 5. DISCUSSIONS AND FUTURE WORK

In this study, we tested the value of curriculum training for classifying Parkinson's disease based on T1-w brain MRI.

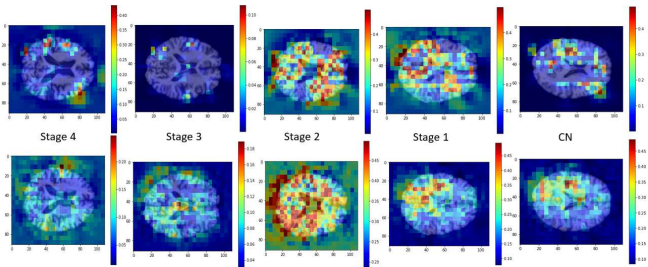

**Fig. 3. Occlusion Sensitivity Analysis**. Here regions of the image are systematically masked with a black patch and the occluded image is then used to create a heatmap based on aggregated CNN predictions. The 'hot' spots indicate regions with the strongest correspondence to the model's prediction. Heatmaps are shown for scans at different H&Y stages. Top: DenseNet121 trained with curriculum learning, Bottom: DenseNet121 trained with regular learning.

To ensure that the tests were not overly optimistic, unlike many prior works we tested on 3 independent datasets from diverse sites around the world, with a range of clinical severity. Overall, and consistent with prior work, classification accuracy was only moderate, with ROC-AUC around 0.65 for the best performing methods. With models that were pre-trained and optimized in a multitask setting, curriculum training appeared to boost performance.

The heat maps generated using the occlusion sensitivity analysis visualize localized regions of the brain that have the strongest correspondence with CNN's predictions. The model trained with curriculum learning achieves better localization of the regions in the image. The earlier stages of PD (stages 1 and 2) show more activity in the heatmaps compared to the higher stages. This may be because effects on the brain differ in earlier stages of PD, compared to the later stages[3].

As classification performance was only modest, future work will add data from other modalities where available (diffusion MRI [22] and DAT-SPECT, as well as quantitative MRI), to determine whether other data sources are more effective. We plan to do experiments to further understand the effect of each additional episode on the final test performance as well as the contribution of the different parts of our proposed loss function. As pre-trained CNNs work well in classifying Alzheimer's disease [9], we also plan to incorporate larger pre-trained models for differential diagnosis of AD versus PD and use multisite data to obtain a realistic assessment of real-world accuracy for these tasks.

## 6. CONCLUSIONS

We show the potential of using a curriculum learning based strategy for Parkinson's disease classification of T1-w MRI scans. The strategy leverages clinically relevant information by progressively increasing the difficulty of the training data. We also used transfer learning and multi-task learning with deep learning models to learn relevant features.

## 7. COMPLIANCE WITH ETHICAL STANDARDS

All datasets were collected with ethics board approval and anonymized and de-identified prior to analysis.

## 8. ACKNOWLEDGMENTS

This work was supported by the U.S. National Institutes of Health, under NIH grant R01NS107513 and U01AG068057.